\documentclass[11pt,a4paper]{article}

\usepackage{amsmath,amssymb}
\usepackage{amsmath}
\usepackage{color}
\usepackage[colorlinks,linkcolor=blue,citecolor=red]{hyperref}
\usepackage{tikz}\usepackage{tikz-cd}
\usetikzlibrary{matrix,arrows}

\DeclareMathSymbol{\bbbr}{\mathalpha}{AMSb}{"52}
\DeclareMathSymbol{\bbbc}{\mathalpha}{AMSb}{"52}

\newcommand{\g}{\mathfrak{g}}
\newcommand{\R}{\mathbb{R}}
\newcommand{\C}{\mathbb{C}}
\newcommand{\La}{\Lambda}
\newcommand{\ot}{\otimes}
\newcommand{\tr}{\operatorname{tr}}
\newcommand\op[1]{\mathop{\rm #1}\nolimits}
\newcommand\E{{\mathcal E}}

\newcommand\com[1]{}
\newcommand\qed{\phantom{\underline{y}}\hfill\hfill$\square$\medskip}

\newcommand\bond[1]{\draw (#1) -- +(1,0)}
\newcommand\lbond[1]{
	\draw (#1) ++(0.03,0.03) -- +(0.94,0);
	\draw (#1) ++(0.03,-0.03) -- +(0.94,0);
	\draw[semithick] (#1) ++(0.45,0) ++(0.15,0.2) -- ++(-0.15,-0.2) -- +(0.15,-0.2)}
\newcommand\tcirc[1]{\filldraw[fill=white,draw=black] (#1) circle (0.08); }
\newcommand\tcross[1]{\draw (#1) ++(-0.12,-0.12)-- +(0.24, 0.24);
	\draw (#1) ++(-0.12, 0.12)-- +(0.24,-0.24); }
\newcommand\wDnode[1]{{\tcirc{#1}}}
\newcommand\xDnode[1]{{\tcross{#1}}}

\newtheorem{theorem}{Theorem}

\newtheorem{lemma}[theorem]{Lemma}

\newtheorem{proposition}[theorem]{Proposition}

\begin{document}

\title{Integrability of dispersionless Hirota type equations \\
and the symplectic Monge-Amp\`ere property}

\author{E.V. Ferapontov$^1$, B. Kruglikov$^2$, V. Novikov$^1$}
     \date{}
     \maketitle
     \vspace{-5mm}
\begin{center}
$^1$Department of Mathematical Sciences \\ Loughborough University \\
Loughborough, Leicestershire LE11 3TU \\ United Kingdom \\
  \bigskip
$^2$Department of Mathematics and Statistics\\
UiT the Arctic University of Norway\\
Tromso 9037, Norway\\
 [2ex]
e-mails: \\[1ex] \texttt{E.V.Ferapontov@lboro.ac.uk}\\
\texttt{Boris.Kruglikov@uit.no} \\
\texttt{V.Novikov@lboro.ac.uk}
\\

\end{center}

\medskip
\begin{abstract}

We prove that integrability of a dispersionless Hirota type equation implies the symplectic
Monge-Amp\`ere property in any dimension  $\geq 4$.
In 4D this 
yields a complete classification of integrable dispersionless PDEs of Hirota type
through a list of heavenly type equations arising in self-dual gravity.
As a by-product of our approach we derive an involutive system of relations characterising symplectic
Monge-Amp\`ere equations in any dimension.

Moreover, we demonstrate that in 4D the requirement of integrability is equivalent to self-duality of
the conformal structure defined by the characteristic variety of the equation on every solution,
which is in turn equivalent to the existence of a dispersionless Lax pair.
We also give a criterion of linerisability of a Hirota type equation
via flatness  of the corresponding conformal structure, and study symmetry properties of integrable equations.

\bigskip

\end{abstract}

\newpage

\tableofcontents

\section{Introduction and the main results}\label{sec:intro}

\subsection{Dispersionless Hirota type equations}\label{sec:Hirota}

Let $u(x^0, \dots, x^{n})$ be a function of $n+1$ independent variables. A dispersionless Hirota type  equation is a scalar second-order PDE for $u$  of the form
 \begin{equation}\label{H}
F(U)=0
 \end{equation}
where $U=Hess (u)=\{u_{\alpha \beta}\}$ is the Hessian matrix of $u$
($u_{\alpha\beta}=\partial_{x^{\alpha}}\partial_{x^{\beta}}u, \ 0\leq\alpha\leq\beta\leq n$).
Equations of type (\ref{H})
appear in a wide range of applications  including the following:

 \begin{itemize}
\item {\bf Integrable systems.} In this context,  Hirota type equations arise as differential relations for  $\tau$-functions of various 3D hierarchies of the dispersionless Kadomtsev-Petviashvili/Toda type, see e.g. \cite{Kodama, Takasaki,  Wiegmann, Zab}.

\item {\bf General relativity.} Symplectic Monge-Amp\`ere equations, which constitute a  subclass of  equations (\ref{H}), are known  to arise  as heavenly type  equations governing self-dual Ricci-flat metrics in 4D \cite{Plebanski, Husain, Grant}.

\item {\bf Differential geometry.} In geometric context equations (\ref{H}), also known as Hessian equations,  appear as relations involving symmetric functions of the eigenvalues of $U$. Their analytical and global aspects  were thoroughly investigated in \cite{Trudinger, Wang, Colesanti, Nadirashvili}, see also references therein.

\item {\bf Submanifolds in Grassmanians.} Equation (\ref{H}) can  be viewed as the defining equation of a hypersurface $X$ in the Lagrangian Grassmannian $\Lambda$, locally parametrised by $(n+1)\times(n+1)$ symmetric matrices $U$. This point of view has been developed in \cite{Fer4, Smith}
leading to remarkable connections with integrable $GL(2, \mathbb{R})$ geometry.
Integrability aspects of 
dispersionless systems related to Grassmann geometries were recently studied in \cite{DFKN2,DFKN3}.
  \end{itemize}

 In what follows we assume that equation (\ref{H}) is {\it non-degenerate} in the sense that the corresponding characteristic variety,
 $$
\sum_{\alpha\leq\beta}\frac{\partial F}{\partial u_{\alpha\beta}}\ p_{\alpha} p_{\beta}=0,
 $$
defines a non-degenerate quadric of rank $n+1$.  This gives rise to the conformal structure
$[g]=g_{\alpha \beta} dx^{\alpha} dx^{\beta}$ where  $(g_{\alpha \beta})$ is the inverse to
the matrix $\Bigl(\frac{1+\delta_{\alpha\beta}}2\frac{\partial F}{\partial u_{\alpha \beta}}\Bigr)$ of the above quadratic form.
It will be demonstrated  that integrability of non-degenerate Hirota type equations 
can be interpreted
in terms of the conformal geometry of $[g]$  (see \cite{Alekseevsky} for  geometry of a special class of degenerate parabolic Monge-Amp\`ere equations).

\subsection{Equivalence group}\label{sec:Sp}

Although we will be primarily interested in the 4D case corresponding to $n=3$, the following properties hold in any dimension. The class of equations (\ref{H}) is invariant under the action of the symplectic group ${\bf Sp}(2n+2,{\bf k})$, where ${\bf k}=\R$ or $\C$ depending on the context.
An element of this group is a block matrix $\begin{pmatrix}A&B\\ C&D\end{pmatrix}$
with $(n+1)\times (n+1)$ matrices $A, B, C, D$ satisfying the defining relations
$A^tC=C^tA$, $B^tD=D^tB$, $A^tD-C^tB=id$, with the action on the Lagrangian Grassmannian $\Lambda$ defined as
 \begin{equation}\label{Sp}
U\mapsto\tilde U=(AU+B)(CU+D)^{-1}.
 \end{equation}
Transformations of this type preserve the integrability, and constitute a natural {\it equivalence group} of the problem.
The corresponding  infinitesimal generators are as follows:
 \begin{gather*}
X_{\alpha \beta}=\frac{\partial}{\partial u_{\alpha \beta}}, \qquad
L_{\alpha \beta}=\sum_{\gamma} u_{\beta \gamma}\frac{\partial}{\partial u_{\alpha \gamma}}+u_{\alpha \beta} \frac{\partial}{\partial u_{\alpha \alpha}},\\
P_{\alpha \beta}=2\sum_{\gamma} u_{\alpha \gamma}u_{\beta \gamma}\frac{\partial}{\partial u_{\gamma \gamma}}+\sum_{\gamma \ne \delta}u_{\alpha \gamma}u_{\beta \delta} \frac{\partial}{\partial u_{\gamma \delta}};
 \end{gather*}
note that $X_{\alpha \beta}=X_{\beta \alpha}$ and $P_{\alpha \beta}= P_{\beta \alpha}$, while  $L_{\alpha \beta}\ne L_{\beta \alpha}$.  Thus, we have
 $(n+1)(n+2)/2$ operators $X_{\alpha \beta}$,  $(n+1)^2$ operators $L_{\alpha \beta}$ and  $(n+1)(n+2)/2$ operators $P_{\alpha \beta}$.
 Altogether, they form the Lie algebra $\mathfrak{sp}(2n+2)$ of dimension $(n+1)(2n+3)$.
Let us represent equation (\ref{H}) in evolutionary form,
\begin{equation}
u_{00}=f(u_{01},\dots,u_{0n},u_{11},u_{12},\dots,u_{nn}).
\label{m0}
\end{equation}
The action  of the equivalence group ${\bf Sp}(2n+2)$ on hypersurfaces in $\Lambda$ induces a (local) action of the same group
(equivalently, its Lie algebra)
on the space
$J^1\bigl(\mathbb{R}^{\frac{n(n+3)}{2}}\bigr)$
of 1-jets of the function $f$ of  variables $u_{0i}, u_{ij}$ ($1\leq i\leq j\leq n$).
This space has dimension $n(n+3)+1$ with coordinates $u_{0i}, u_{ij}, f,  f_{u_{0i}}, f_{u_{ij}}$.

It is easy to see that the induced action  has a unique Zariski open orbit (its complement consists of 1-jets of degenerate systems). This property allows one to assume that all sporadic factors depending on first-order derivatives of $f$  that arise in the process of Gaussian elimination in the proofs of our main results in Section \ref{sec:Proofs} are nonzero. This considerably simplifies the arguments by eliminating unessential branching.
Furthermore, in the verification of various polynomial identities involving higher-order partial derivatives of $f$ one can, without any loss of generality, give the first-order derivatives of $f$ any numerical values
corresponding to a non-degenerate 1-jet: this often renders otherwise impossible computations manageable.

\subsection{Integrability by the method of hydrodynamic reductions}\label{sec:int}

Integrability of Hirota type equations (\ref{H}) can be approached based on the method of hydrodynamic reductions \cite{GibTsar, GibTsar1, Fer1, Fer2,  Fer3, Fer4}. In the most general set-up (for definiteness, we restrict to the 4D case), it applies to quasilinear systems of the form
 \begin{equation}
A_0 ({\bf v}){\bf v}_{x^0}+A_1({\bf v}){\bf v}_{x^1}+A_2({\bf v}){\bf v}_{x^2}+A_3({\bf v}){\bf v}_{x^3}=0,
\label{quasi1}
 \end{equation}
where ${\bf v}=(v^1, ..., v^m)^t$ is an $m$-component column vector of the dependent variables $x^{\alpha}$ and $A_{\alpha}$ are $l\times m$ matrices where the number $l$ of the equations is allowed to exceed the number $m$ of the unknowns. Note that equation (\ref{H}) can be cast into  quasilinear form  (\ref{quasi1}) by choosing $u_{\alpha \beta}$ as the new dependent variables ${\bf v}$ and writing out the compatibility conditions among them, see Section \ref{sec:Proofs}. The method of hydrodynamic reductions consists of seeking multi-phase solutions in the form
 \begin{equation}
{\bf v}={\bf v}(R^1, ..., R^N)
 \label{ansatz}
 \end{equation}
where the phases $R^I({\bf x})$, whose number $N$ is allowed to be arbitrary, are required to satisfy a triple of consistent $(1+1)$-dimensional systems,
 \begin{equation}
R^I_{x^2}=\mu^I(R) R^I_{x^1}, ~~~ R^I_{x^3}=\nu^I(R) R^I_{x^1}, ~~~ R^I_{x^0}=\lambda^I(R) R^I_{x^1},
\label{R}
 \end{equation}
known as systems of hydrodynamic type. The corresponding characteristic speeds must satisfy the commutativity conditions \cite{Tsar, Tsar1},
 \begin{equation}
\frac{\partial_J\mu^I}{\mu^J-\mu^I}=\frac{\partial_J\nu^I}{\nu^J-\nu^I}=\frac{\partial_J\lambda^I}{\lambda^J-\lambda^I},
\label{comm}
 \end{equation}
here $I\ne J, \  \partial_J=\partial_{ R^J}, \ I, J=1, \dots, N$.
Equations (\ref{R}) are said to define an $N$-component  hydrodynamic reduction of  system (\ref{quasi1}). System (\ref{quasi1}) is said to be {\it integrable} if, for every $N$, it possesses infinitely many $N$-component hydrodynamic reductions parametrised by $2N$ arbitrary functions of one variable \cite{Fer1, Fer3}. This requirement imposes strong constraints (integrability conditions) on the matrix elements of  $A_{\alpha}({\bf v})$.

The method of hydrodynamic reductions has been successfully applied to the class of 3D Hirota type equations, leading to extensive classification results and remarkable geometric relations \cite{Fer4}.
In the present paper we directly apply the method  to  the class of 4D Hirota type equations. The 4D situation turns out to be far more restrictive, in particular, we demonstrate that  the requirement of integrability  implies the symplectic Monge-Amp\`ere property.

\subsection{Symplectic Monge-Amp\`ere equations}\label{sec:symp}

A symplectic Monge-Amp\`ere equation is obtained by equating to zero a linear combination of  minors  (of all possible orders) of the Hessian matrix $U=Hess(u)$. These equations constitute a proper subclass of Hirota type equations (\ref{H}). Geometrically, the corresponding hypersurface $X\subset \Lambda$ is a hyperplane section  of the Pl\"ucker embedding of the Lagrangian Grassmannian $\Lambda$.  Among the most well-studied examples one should primarily mention the equations
$$
\det U=\tr U ~~~ {\rm and} ~~~ \det U=1,
$$
governing special Lagrangian submanifolds and affine hyperspheres, respectively \cite{Joyce, Calabi}
(both non-integrable for $n\geq 2$).

In 2D, any symplectic Monge-Amp\`ere equation is linearisable \cite{Kushner}.
In 3D,  integrability of a  symplectic Monge-Amp\`ere equation is equivalent to its  linearisability \cite{Fer4}.
In 4D, non-degenerate integrable symplectic Monge-Amp\`ere equations were  classified in \cite{DF}:

 \begin{theorem}\label{SMA}
Over  the field of complex numbers, any $4D$ integrable non-degenerate symplectic
Monge-Amp\`ere equation   is ${\bf Sp}(8)$-equivalent to one of the 6  normal
forms:
\begin{enumerate}
\item $u_{00}-u_{11}-u_{22}-u_{33}=0$ (linear wave equation);
\item $u_{02}+u_{13}+u_{00}u_{11}-u_{01}^2=0$ (second heavenly equation \cite{Plebanski});
\item $u_{02} - u_{01}u_{33}+u_{03}u_{13}=0$ (modified heavenly equation \cite{DF});
\item $u_{02}u_{13}-u_{03}u_{12}-1=0$ (first heavenly equation \cite{Plebanski});
\item $u_{00}+u_{11}+u_{02}u_{13}-u_{03}u_{12}=0$ (Husain equation \cite{Husain});
\item $\alpha u_{01}u_{23} + \beta u_{02}u_{13}+\gamma u_{03}u_{12} = 0$ (general
heavenly equation \cite{Schief}), $\alpha+\beta+\gamma=0$.
\end{enumerate}
 \end{theorem}

Equations 2-6 are known to be non-linearisable, and contact non-equivalent.  All of them originate from self-dual Ricci-flat geometry, and have been thoroughly investigated in the literature. Thus, bi-Hamiltonian formulation of heavenly type equations was established in \cite{ Nutku, Nutku1,
Sheftel1, Sheftel2}. Twistor-theoretic aspects of the associated hierarchies were discussed in \cite{Takasaki1, Dun1, Dun2, Bogdanov1, Bogdanov3}. The integrability by the method of hydrodynamic reductions was demonstrated  in \cite{Fer2, Fer3}. Symmetries and recursion operators  were constructed in \cite{Takasaki2, Strachan1, Sheftel, KrugMor1, KrugMor2, pryk, MarSer}.  A $\bar \partial$-approach and a novel version of the inverse scattering transform  were developed in \cite{Bogdanov2, ManSan1, ManSan2}.


It was conjectured in \cite{DF} that in 4D, the requirement of integrability of equation (\ref{H}) implies the symplectic Monge-Amp\`ere property. The proof of this conjecture, which is the main result of our paper,  is given in Section \ref{sec:Proofs}. Together with Theorem \ref{SMA}, this completes the classification of integrable Hirota type equations in 4D.

The Monge-Amp\`ere property comes as the result of a rather challenging calculation: starting with evolutionary form (\ref{m0}), we derive the integrability conditions, which constitute complicated differential relations that are linear in the third-order, and quadratic in the second-order partial derivatives of $f$.
In 3D, these relations can be uniquely solved for all third-order partial derivatives of $f$ resulting in an involutive system of integrability conditions \cite{Fer4}. The first remarkable phenomenon of the 4D case is the appearance, along with third-order relations, of a whole set of additional  second-order relations that are quadratic in the second-order partial derivatives of $f$. The second remarkable phenomenon is that the ideal generated by these quadratic relations possesses a  linear radical responsible for the Monge-Amp\`ere property.

To establish the Monge-Amp\`ere property we need the corresponding  set of differential constraints for  $f$.
These have only been known  in low dimensions \cite{Ruggeri,Colin, Gutt}.
In Section \ref{sec:MA} we derive these constraints in any dimension by using formal theory
of differential equations and representation theory.

 \begin{theorem}\label{T0}
Equation (\ref{m0}) is of symplectic Monge-Amp\`ere type if and only if $d^2f$ is a linear combination of the second fundamental forms of the Pl\"ucker embedding of the Lagrangian Grassmannian $\Lambda$ restricted to the hypersurface defined by (\ref{m0}).
This property is characterised by $N(n)=\frac1{24}n(n+1)(n+2)(n+7)$ relations
\eqref{E1ab}-\eqref{E4} from Section \ref{sec:MA}
which are second-order quasilinear PDEs for  $f$.
 \end{theorem}

In Section \ref{sec:MA} we provide a characterisation of symplectic Monge-Amp\`ere equations represented in
implicit (non-evolutionary) form \eqref{sMAEs} by an alternative set of linear differential constraints, see Theorem \ref{xMAE}.

\subsection{Integrability, self-duality and Lax pairs}\label{conf}

In 4D, the key  invariant of a conformal structure  $[g]$ is its Weyl tensor $W$.
A conformal structure is said to be self-dual  if,  with a proper choice of orientation,
 \begin{equation*}
W=* W,
 \end{equation*}
where $*$ is the Hodge star operator. Integrability of the conditions of self-duality  by the twistor construction is due to Penrose \cite{Penrose}
who observed that self-duality of $[g]$ is equivalent to the existence of a 3-parameter family of totally null surfaces ($\alpha$-surfaces).
 In Section \ref{sec:Proofs} we prove that integrability of a 4D equation (\ref{H}) is equivalent to the requirement that the conformal structure $[g]$ defined by the characteristic variety of the equation
must be self-dual on every solution. Thus, for the second heavenly equation we have
 $$
g=dx^0dx^2+dx^1dx^3-u_{11}(dx^2)^2+2u_{01}dx^2dx^3-u_{00}(dx^3)^2,
 $$
and a direct calculation shows that the conformal structure $[g]$ is indeed self-dual on every solution.
Summarising, {\it solutions to integrable systems carry integrable conformal geometry}.

It is known that all equations from Theorem \ref{SMA} possess dispersionless Lax pairs, that is, there exist vector fields $X, Y$ depending on $u_{\alpha \beta}$ and an auxiliary parameter $\lambda$ such that the commutativity condition $[X, Y]=0$ holds identically modulo the equation (and its differential consequences).  For the second heavenly equation we have
$$
X=\partial_3+u_{11}\partial_1-u_{01}\partial_0+\lambda \partial_0, ~~~
Y=\partial_2-u_{01}\partial_1+u_{11}\partial_0-\lambda \partial_1,
$$
here $\partial_{\alpha}=\frac{\partial}{\partial x^{\alpha}}$.
Integral surfaces of the involutive distribution $\langle X, Y\rangle$
provide  null surfaces of the corresponding conformal structure $[g]$, thus establishing its self-duality.

More generally, for Hirota type equations (\ref{H}), disperionless integrability
(i.e.\ integrability by the method of hydrodynamic reductions)
is equivalent to the existence of a Lax pair in commuting vector fields. Due to the characteristic property of Lax pairs established in \cite{CalKrug}, this implies self-duality of the corresponding conformal structure $[g]$ on every solution. In fact, one can say more: the absence of $\partial_{\lambda}$ in the vector fields $X, Y$ defining the Lax pair (which is the case for all integrable  Monge-Amp\`ere equations from Theorem \ref{SMA}) implies that the corresponding conformal structure $[g]$ is   hyper-Hermitian  \cite{Dun5}.

\subsection{Hirota type equations: summary of  results}\label{sec:sum}

In 4D we obtain a complete classification of integrable non-degenerate Hirota type equations:

 \begin{theorem}\label{T1}
For non-degenerate equations (\ref{H}) in 4D, the following conditions are equivalent:
\medskip

\noindent (a) Equation (\ref{H}) is integrable by the method of hydrodynamic reductions.\medskip

\noindent (b) Conformal structure $[g]$ defined by the characteristic variety of equation (\ref{H}) is self-dual on every solution (upon complexification, or in real signatures (4,0), (2,2)).\medskip

\noindent (c) Equation (\ref{H}) possesses a
dispersionless Lax pair.\medskip

\noindent (d) Equation (\ref{H})  is ${\bf Sp}(8)$-equivalent (over $\C$) to one of the 6  canonical forms of integrable symplectic Monge-Amp\`ere equations classified in \cite{DF}.
 \end{theorem}

The proof of Theorem \ref{T1} is given in Section \ref{sec:Proof-2}.
As a corollary, we obtain  the following characterisation of linearisable equations (Section \ref{sec:linearisation}):

 \begin{theorem}\label{T2}
A non-degenerate Hirota type equation in 4D is linearisable by a transformation from the equivalence group ${\bf Sp}(8)$
if and only if the associated conformal structure $[g]$ is flat on every solution (this statement is true in  both  real and complex situations).
 \end{theorem}


In addition, we investigate symmetry aspects of integrability in 4D.
As noted in \cite{DF}, every four-dimensional integrable symplectic Monge-Amp\`ere equation is
invariant under a subgroup of the equivalence group ${\bf Sp}(8)$, of dimension at least 12.
As another corollary of Theorem \ref{T1} we obtain the following result (Section \ref{symalg}):

 \begin{theorem}\label{T3}
Let $X^9\subset\Lambda^{10}$ be a hypersurface in the Lagrangian Grassmannian corresponding to an integrable Hirota type equation in 4D. 
Then it is almost homogeneous:  there exists a subgroup of the equivalence group that acts on $X^9$ with a Zariski open orbit.
 \end{theorem}

The fact that every integrable Hirota type equation in 4D possesses nontrivial symmetries from the equivalence group
${\bf Sp}(8)$ (as well as many more from the general contact group) is in sharp contrast with
the situation in 3D \cite{Fer4} where a generic integrable Hirota type equation was shown to possess
no continuous symmetries from the equivalence group ${\bf Sp}(6)$.
We also note that for the two-component first-order systems in 4D studied in \cite{DFKN3} the integrability
implies a certain amount of symmetry from the corresponding equivalence group,
yet in general this does not make the equation $X$ almost homogeneous.


Finally, we obtain the following generalisation of one of the statements from Theorem \ref{T1}.
According to it, any integrable Hirota type equation in 4D is necessarily
of Monge-Amp\`ere type (in 3D this is not true, see \cite{Fer4}).
This property persists in higher dimensions (Section \ref{final-higherD}):
 \begin{theorem}\label{T4}
 In all dimensions higher than 3, the integrability of a non-degenerate Hirota type equation implies the symplectic Monge-Amp\`ere property.
 \end{theorem}
 

In particular, in higher dimensions $n+1\ge4$, the local integrability constraints on a smooth
hypersurface $X\subset\Lambda$ imply its global algebraicity.

\section{Characterisation of symplectic Monge-Amp\`ere equations}\label{sec:MA}

In this section we  consider Hirota type  equations represented in evolutionary form (\ref{m0}). The proof of Theorem \ref{T1} will require differential constraints for the right-hand side $f$  that characterise symplectic Monge-Amp\`ere equations. In dimensions $\leq 4$ these were obtained in \cite{Boillat, Ruggeri, Colin, Gutt} based on linear degeneracy (complete exceptionality) of Monge-Amp\`ere equations.
Here we complete the case of general dimension which was left open.

We  adopt a differential-geometric point of view that identifies equation (\ref{m0}) with a hypersurface $X$ in the Lagrangian Grassmannian $\Lambda$ \cite{Fer4, DF}.  The Monge-Amp\`ere property is equivalent to the requirement that  osculating spaces to $X$ span a hyperplane in the projective space of the Pl\"ucker embedding of $\Lambda$. The last condition can be  represented as a  simple relation among the second fundamental forms of $X$, leading to the required differential constraints. The presentation below  follows \cite{DFKN1}, see also \cite{Gutt}.

Once the constraints characterising the symplectic Monge-Amp\`ere property are derived we show their
completeness by demonstrating involutivity of the corresponding system of PDEs. Theorem \ref{T0} then
follows from  Section \ref{sec:MAh}. We also derive, in Section \ref{sec:MAEs}, a linear system characterising the
symplectic Monge-Amp\`ere property in implicit form.

\subsection{Monge-Amp\`ere equations in 2D}\label{sec:MA2}

Let us begin with the two-dimensional  situation which however contains all essential
ingredients of the general case. Setting $r=u_{00},\ s=u_{01},\ t=u_{11}$
we represent  equation (\ref{m0}) as
 \begin{equation}\label{m1}
r=f(s,t).
 \end{equation}
 Equation (\ref{m1}) specifies a surface $X^2$ in the Lagrangian Grassmannian $\Lambda^3$ which is identified with $2\times 2$ symmetric matrices,
 $$
U=\left(\begin{array}{cc} r & s \\ s & t\end{array}\right).
 $$

 \begin{proposition}\label{prop1}
Equation (\ref{m1}) is of Monge-Amp\`ere type if and only if the symmetric differential $d^2f$
is proportional to the quadratic form $drdt-ds^2$ restricted to $X^2$:
\begin{equation}
d^2f \in {\rm span}  \langle dfdt-ds^2\rangle.
\label{m2}
\end{equation}
 \end{proposition}

\noindent {\bf Proof:} The Pl\"ucker embedding $\Lambda^3\hookrightarrow \mathbb{P}^4$  is a quadric with
position vector $(t,\,s,\,r,\, tr-s^2)$. The induced embedding  of $X^2$ has position vector
 $$
R=(t,\, s,\, f,\, tf-s^2).
 $$
To prove that equation (\ref{m1}) is of Monge-Amp\`ere type we need to show that  components of $R$ satisfy a linear relation with constant coefficients or, equivalently, that the Pl\"ucker image of $X^2$
belongs to a hyperplane in $\mathbb{P}^4$. This means that the union of all osculating spaces to $X^2$
must be 3-dimensional. Since the tangent space of $X^2$, which is spanned by the vectors
 $$
R_s=(0,\, 1,\, f_s,\, tf_s-2s),\qquad R_t=(1,\, 0,\, f_t,\, tf_t+f),
 $$
is already 2-dimensional, we have to show that the union of the second and third-order osculating spaces
(spanned by the second and third-order partial derivatives of the position vector $R$ with respect to $s,t$),
is only 1-dimensional. As higher-order derivatives of $R$ have zeros in the first two positions,
we obtain the following rank condition:
 $$
\op{rank}
\left(\begin{array}{ll}
f_{ss} &   tf_{ss}-2\\
f_{st} &   tf_{st}+f_{s}\\
f_{tt} &   tf_{tt}+2f_{t}\\
f_{sss} &   tf_{sss}\\
f_{sst} &   tf_{sst}+f_{ss}\\
f_{stt} &   tf_{stt}+2f_{st}\\
f_{ttt} &   tf_{ttt}+3f_{tt}
\end{array}\right)
=\op{rank}
\left(\begin{array}{lr}
f_{ss} &   -2\\
f_{st} &   f_{s}\\
f_{tt} &   2f_{t}\\
f_{sss} & 0\\
f_{sst} & f_{ss}\\
f_{stt} &  2f_{st}\\
f_{ttt} &  3f_{tt}
\end{array}\right)
=1.
 $$
In other words, the second column of the last matrix should be proportional to the first one.
Denoting by $p$ the proportionality coefficient, this can be represented in  compact form as
 \begin{gather}
d^2f=2p\cdot(dfdt-ds^2),\label{m3}\\
d^3f=3p\cdot dt\,d^2f.\label{m4}
 \end{gather}
Calculating the (symmetric) differential of (\ref{m3}) and comparing the result with  (\ref{m4}) we obtain
the following equation for $p$:
 \begin{equation}
dp=p^2dt.\label{m5}
 \end{equation}
Equations (\ref{m3}) and (\ref{m5}) constitute a closed involutive differential system for $f$ which characterises symplectic Monge-Amp\`ere equations in 2D. It remains to point out that
(\ref{m5}) can be obtained from the consistency conditions of equations (\ref{m3}), without invoking (\ref{m4}). In other words, equations (\ref{m3}) imply both (\ref{m4}) and (\ref{m5}).
This finishes the proof of Proposition \ref{prop1}. \qed

\noindent {\bf Remark:} Proposition \ref{prop1} has a clear projective-geometric interpretation.
The second fundamental forms of the surface $X^2\subset \Lambda^3 \subset \mathbb{P}^4$
are spanned by $d^2f$ and $dfdt-ds^2$. Here the last form is the restriction to $X^2$ of the second fundamental form of the Grassmannian $\Lambda^3$ itself, namely, $drdt-ds^2$.
The claim is thus that the only `essential' second fundamental form of $X^2$ is the one coming from the second fundamental form of  $\Lambda^3$. This property is clearly a necessary  condition for $X^2$ to be a hyperplane section, and the above proof shows  that it is also sufficient.

\medskip

\noindent {\bf Remark:} Elimination of $p$ from (\ref{m3}) leads to the following system of PDEs for $f$:
 \begin{equation}\label{1index}
f_t f_{ss}+f_{tt}=0, ~~~ f_s f_{ss}+2f_{st}=0.
 \end{equation}
The general solution of this system is given by the formula
 $$
f=\frac{s^2-\beta s-\gamma t-\delta}{t+\alpha},
 $$
which indeed specifies a  Monge-Amp\`ere equation, $rt-s^2+\alpha r+\beta s+\gamma t+\delta=0$.


\subsection {Monge-Amp\`ere equations in higher dimensions: the defining relations}\label{sec:MA3}

A 3D equation (\ref{m0})$_{n=2}$ specifies a hypersurface $X^5$ in the  Lagrangian Grassmannian $\Lambda^6$. We will work in the  affine chart of $\Lambda^6$  identified with the space of $3\times 3$
symmetric matrices
 $$
U=\left(
\begin{array}{ccc}
u_{00} & u_{01} & u_{02}\\
u_{01} & u_{11} & u_{12}\\
u_{02} & u_{12} & u_{22}\\
\end{array}
\right).
 $$
The minors of $U$ define the Pl\"ucker embedding $\Lambda^6  \hookrightarrow \mathbb{P}^{13}$. The second fundamental forms of this embedding  are given by   $2\times 2$ minors of the matrix $dU$. Since the second osculating space of $\Lambda^6\subset \mathbb{P}^{13}$ is only $12$-dimensional, there is also a third fundamental form, namely $\det  dU$.

For general $n$ equation (\ref{m0}) defines a hypersurface $X$ in the  Lagrangian Grassmannian
$\Lambda$ of dimension $d(n+1)$, where $d(n)=\frac{n(n+1)}{2}$ is the number
of entries of a symmetric $n\times n$ matrix. The Lagrangian Grassmannian is
embedded via the Pl\"ucker map into projective space of dimension $p(n+1)-1$, where
$p(n)=\frac{2(2n+1)!}{n!(n+2)!}$ is the number of independent minors of a symmetric $n\times n$
matrix:
\smallskip
 \begin{equation}\label{X2L}
X^{d(n+1)-1}\subset\Lambda\hookrightarrow \mathbb{P}^{p(n+1)-1}.
 \end{equation}

 \begin{proposition} \label{prop2}
Equation (\ref{m0}) is of Monge-Amp\`ere type if and only if $d^2f$ belongs to the span of the second fundamental forms of the Pl\"ucker embedding of $\Lambda$ restricted to the hypersurface $X$.
 \end{proposition}

We will prove this fully in the 3D case, the general case can be proved similarly for successive
dimensions $n$. The condition of the proposition is clearly necessary,  for general $n$ the sufficiency
will follow from the theorem in the next section.

\medskip

\noindent {\bf Proof:} For $n=2$ we have to require  that  the induced Pl\"ucker embedding of $X^5$
belongs to a hyperplane in $\mathbb{P}^{13}$, i.e.\ the union of all osculating spaces to $X^5$ is 12-dimensional. Calculations analogous to that from Section \ref{sec:MA2} yield the following expansions
of the differentials of $f$:
 \begin{equation}\label{m31}
\begin{array}{c}
d^2f=2a_0(dfdu_{12}-du_{01}du_{02})+2a_1(dfdu_{22}-(du_{02})^2)+2a_2(dfdu_{11}-(du_{01})^2)
\vphantom{\frac{A_A}{A^a_A}}\\
+2b_0(du_{11}du_{22}-(du_{12})^2)+2b_1(du_{01}du_{12}-du_{02}du_{11})+2b_2(du_{02}du_{12}-du_{01}du_{22}),
\end{array}
 \end{equation}
 \begin{equation}\label{m41}
d^3f=3\omega d^2f+6s \det dU\vert_{u_{00}=f},
 \end{equation}
where $\omega=a_0du_{12}+a_1du_{22}+a_2du_{11}$.
Compatibility conditions for the relations (\ref{m31})-(\ref{m41}) implies:
 \begin{equation}\label{m51}
\begin{array}{c}
da_0=a_0\omega-2sdu_{12}, ~~~ da_1=a_1\omega+sdu_{11}, ~~~da_2=a_2\omega+sdu_{22}, \\
db_0=b_0\omega+sdf, ~~~ db_1=b_1\omega+{2} s du_{02}, ~~~db_2=b_2\omega+{2}sdu_{01}, ~~~  ds=s\omega.
\end{array}
 \end{equation}
One can verify that $d\omega =0$. Equations (\ref{m31}) and (\ref{m51}) constitute an  involutive differential system for $f$  which characterises Monge-Amp\`ere equations. It remains to point out that  equations (\ref{m51}) can be obtained from the consistency of  equations (\ref{m31}) alone, without invoking (\ref{m41}). In other words, equations (\ref{m31}) imply both (\ref{m41}) and (\ref{m51}). This finishes the proof of Proposition \ref{prop2}. \qed

Proposition \ref{prop2} leads to a system of PDEs for $f$ (for $n=2$ these can be obtained by eliminating coefficients $a_i, b_i$ from equations (\ref{m31})).
First of all, for every index $i=1,\dots,n$ one has the analogues of (\ref{1index}),
 \begin{equation}\label{E1ab}
f_{u_{ii}}f_{u_{0i}u_{0i}}+f_{u_{ii}u_{ii}}=0, ~~~ f_{u_{0i}}f_{u_{0i}u_{0i}}+2f_{u_{0i}u_{ii}}=0.
 \end{equation}
Secondly, for every pair of indices $i\ne j\in\{1,\dots,n\}$ one has the relations
 \begin{gather}
f_{u_{0j}}f_{u_{0i}u_{0i}}+2f_{u_{0i}}f_{u_{0i}u_{0j}}+2f_{u_{0i}u_{ij}}+2f_{u_{0j}u_{ii}}=0,\label{E2a}\\
f_{u_{ij}}f_{u_{0i}u_{0i}}+2f_{u_{ii}}f_{u_{0i}u_{0j}}+2f_{u_{ii}u_{ij}}=0,\label{E2b}\\
f_{u_{jj}}f_{u_{0i}u_{0i}}+f_{u_{ii}}f_{u_{0j}u_{0j}}+2f_{u_{ij}}f_{u_{0i}u_{0j}}+2f_{u_{ii}u_{jj}}+f_{u_{ij}u_{ij}}=0.\label{E2c}
 \end{gather}
Based on a different approach, for $n=2$ these relations  were derived in \cite{Ruggeri}, see also \cite{Gutt}.
Futhermore, for every triple of  
distinct indices $i\ne j\ne k \in\{1,\dots,n\}$ one has the relations
 \begin{gather}
f_{u_{0k}}f_{u_{0i}u_{0j}}+f_{u_{0j}}f_{u_{0i}u_{0k}}+f_{u_{0i}}f_{u_{0j}u_{0k}}+f_{u_{0i}u_{jk}}+f_{u_{0j}u_{ik}}+f_{u_{0k}u_{ij}}=0,\label{E3a}\\
f_{u_{jk}}f_{u_{0i}u_{0i}}+2f_{u_{ik}}f_{u_{0i}u_{0j}}+2f_{u_{ij}}f_{u_{0i}u_{0k}}+2f_{u_{ii}}f_{u_{0j}u_{0k}}+2f_{u_{ii}u_{jk}}+2f_{u_{ij}u_{ik}}=0.
\label{E3b}
 \end{gather}
For $n=3$ the above relations \eqref{E1ab}-\eqref{E3b} were obtained in \cite{Colin}
(25 relations altogether).
Finally, for every four distinct indices $i\ne j\ne k \ne l \in\{1, \dots, n\}$ one has the relations
 \begin{multline}\label{E4}
f_{u_{kl}}f_{u_{0i}u_{0j}}+f_{u_{jl}}f_{u_{0i}u_{0k}}+f_{u_{jk}}f_{u_{0i}u_{0l}}+f_{u_{il}}f_{u_{0j}u_{0k}}+f_{u_{ik}}f_{u_{0j}u_{0l}}+f_{u_{ij}}f_{u_{0k}u_{0l}}\\
+f_{u_{ij}u_{kl}}+f_{u_{ik}u_{jl}}+f_{u_{il}u_{jk}}=0.
 \end{multline}

\noindent {\bf Remark:}  Relations (\ref{E1ab})-(\ref{E4}) can be obtained from relations (\ref{1index}) via a traveling wave reduction.  Let $u(x^0, \dots, x^n)$ solve a Monge-Amp\`ere equation represented in evolutionary form (\ref{m0}). Consider the traveling wave ansatz $u(x^0, \dots, x^n)=u(x^0, \xi)$ where
$$
\xi=\alpha_1x^1+\dots +\alpha_nx^n, ~~~ \alpha_i=const.
$$
Using $u_{0i}=\alpha_iu_{0\xi}, \ u_{ij}=\alpha_i\alpha_ju_{\xi \xi}$ we can reduce (\ref{m0}) to a 2D Monge-Amp\`ere equation of type (\ref{m1}). Imposing relations (\ref{1index}) we obtain expressions that are quartic  in $\alpha$'s. Equating similar terms we recover all  relations (\ref{E1ab})-(\ref{E4}).

\subsection{Monge-Amp\`ere equations in evolutionary form: proof of Theorem \ref{T0}}\label{sec:MAh}


In this section we apply the Spencer machinery \cite{Sp} to show that 
the system of relations (\ref{E1ab})-(\ref{E4}) defines the class of Monge-Amp\`ere equations.
Note that every differential system can be described  by either its defining relations (PDEs) or jets of its solutions.
In the latter case the system is involutive, but we do not have control over the defining relations.
In the former case we have control but do not know compatibility a priori. To demonstrate that these two descriptions coincide
we first prove that Monge-Amp\`ere equations have defining
relations of the second order only, then by dimensional reasons we conclude that these must coincide with
relations (\ref{E1ab})-(\ref{E4}).

 \begin{theorem}
Hirota type equation \eqref{m0} is of Monge-Amp\`ere type if and only if the right hand side $f$ satisfies relations (\ref{E1ab})-(\ref{E4}).
 \end{theorem}

\smallskip

\noindent {\bf Proof.}  Let $\E\subset J^\infty(\R^{\bar d})$ denote the system of PDEs for $f$ characterising the
Monge-Amp\`ere property. Here ${\bar d}=d(n+1)-1$, $\R^{\bar d}$ is the space of independent
arguments of $f$ and  $\E_0=J^0=\R^{\bar d+1}$ is an open chart in $\Lambda$ (here $d(n+1)$ and $p(n+1)$  are the same as in Section \ref{sec:MA3}).
Locally, we make the identification $X=\op{graph}(f)\subset J^0$.
Referring to \cite{Kras,KL} for the basics of jet-theory and the formal theory of PDEs,
we identify the system $\E$ with a co-filtered subset in jets, meaning that the sequence
$\E_k\subset J^k(X)$ forms a tower of bundles $\pi_{k,l}:\E_k\to\E_l$ for $k>l$.
Clearly $\E_1=J^1$. We want to prove  that $\E_2$ is generated by a system of
relations $\psi_j=0$ on 2-jets given by \eqref{E1ab}-\eqref{E4}.
It is however easier to study the properties of $\E$ by looking at its solutions.

Thus we identify $\E_k$ as the space of $k$-jets of the set of all Monge-Amp\`ere equations
written in the evolutionary form \eqref{m0}. By definition the symbol spaces of $\E$  are
the subspaces $g_k=\op{Ker}(d\pi_{k,k-1}:T\E_k\to T\E_{k-1})\subset S^k\tau^*$ where
$\tau=T_oX$ is the model tangent space. For $k=0,1$ we have: $g_0=\R$, $g_1=\tau^*$.
For $k\ge2$ the symbols $g_k$ can be interpreted in terms of the
Hessian matrix $U$. Indeed, any Monge-Amp\`ere equation is a relation of the form
 \begin{equation}\label{eqM}
M_0+M_1+\dots+M_n+M_{n+1}=0,
 \end{equation}
where $M_i$ is a linear combination of $i\times i$ minors of $U$.
Linearising this at the identity matrix $I$, i.e.\ setting $U=I+\epsilon A$ and
truncating the higher-order terms in $\epsilon$, the symbol $g_k$ for $k\ge2$ can be interpreted as
the space generated by linearly independent minors of $A$ of size $k$.

It was noted in \cite{MorMathOverflow} that the number of independent $k\times k$ minors of a symmetric
$n\times n$ matrix is $b(k,n)=\frac1{k+1}\binom{n}{k}\binom{n+1}{k}$. The discussion above
implies that $\dim g_k=b(k,n+1)$ for all $k\in[0,n+1]$ with the exception of $k=1$, in which case
$\dim g_1=b(1,n+1)-1={\bar d}$ due to the relation $u_{00}=f$.
For $k>n+1$ the symbol vanishes, $g_k=0$, signifying that the system $\E$ is of finite type.
Its solution space $S$, which can be identified with the dual projective space to $\mathbb{P}^{p(n+1)-1}$
from \eqref{X2L}, is therefore finite-dimensional, and
 $$
\dim S=\sum_{k=0}^\infty\dim g_k=\sum_{k=0}^{n+1}b(k,n+1)-1=p(n+1)-1.
 $$
We claim that $\E_2$ is precisely the locus of relations \eqref{E1ab}-\eqref{E4}.
These relations are independent and vanish on every Monge-Amp\`ere equation.
Thus $\{\psi_j\}$ contain the derived equations. On the other hand, their count is
as follows: $n+n$ relations \eqref{E1ab}, $n(n-1)$ relations \eqref{E2a}, $n(n-1)$ relations \eqref{E2b},
$\binom{n}2=\frac{n(n-1)}2$ relations \eqref{E2c}, $\binom{n}3$ relations \eqref{E3a},
$3\binom{n}3=\frac{n(n-1)(n-2)}2$ relations \eqref{E3b}, and
$\binom{n}4$ relations \eqref{E4}. These numbers sum up to $N(n)=\binom{d(n+1)}2-b(2,n+1)$
which is the codimension of $g_2\subset S^2T^*X$.
This count along with the  quasi-linearity of relations $\psi_j$ implies the claim.

To finish the proof we observe that the higher symbols $g_{k+2}$ for $k>0$ coincide with the prolongations
$g_2^{(k)}:=S^{k+2}\tau^*\cap S^k\tau^*\otimes g_2$, this is the statement of Lemma \ref{LeM} below.
Thus the number of relations specifying the prolongation
$\E_2^{(k)}$ (the locus of the prolonged equations $D_\sigma\psi_j=0$ for all multi-indices
$\sigma$ of length $|\sigma|\leq k$) is no less than that for $\E_{k+2}$. But it cannot be bigger
because otherwise the solution space of the prolonged system $\E_2$ will be less than that of
$\E$ (which contains all Monge-Amp\`ere equations). Thus $\E_2^{(k)}=\E_{k+2}$,
hence the system $\E_2$ given by relations \eqref{E1ab}-\eqref{E4} is formally integrable,
and hence locally solvable for any admissible Cauchy data due to its finite type.
\qed

To justify the above proof it remains to compute prolongations of the symbols used above.
For this we exploit the subalgebra $A_n=\mathfrak{sl}_{n+1}$ in the Lie algebra $C_{n+1}=\g$
of the equivalence group $G={\bf Sp}(2n+2,\C)$: in the $|1|$-grading $\g=\g_{-1}\oplus\g_0\oplus\g_1$
corresponding to the parabolic subalgebra $\mathfrak{p}=\mathfrak{p}_{n+1}$
(numeration: the last node on the Dynkin diagram of $C_{n+1}$ crossed) we have $\g_0=\mathfrak{gl}_{n+1}=\mathfrak{sl}_{n+1}\oplus\R$ and this naturally acts
on the tangent space to the Lagrangian Grassmannian $\Lambda=G/P$ (with $\mathfrak{p}=\op{Lie}(P)$).
Thus the tangent and symbol spaces are
all $A_n$-modules. Below $\Gamma_\mu$ indicates the irreducible $A_n$-representation with the highest weight $\mu$ that we decompose by the fundamental weights $\lambda_i$.

Denote by $\hat{\E}$ the equation-manifold describing Monge-Amp\`ere equations in
implicit form \eqref{eqM}.
Interpreted in jet-formalism (similar to the above proof) as a tower of bundles $\hat{\E}_k$,
we obtain their symbol spaces $\hat{g}_k=\op{Ker}(\pi_{k,k-1}:\hat{\E}_k\to\hat{\E}_{k-1})$;
note that we do not need to pass to tangent spaces as the equation $\hat{\E}$
(as well as its solution spaces) is linear.
Since the equation $\hat{\E}$ is specified by its solutions, it is involutive.
However, a-priori, it can have PDE-generators of different orders. We will show that this is not the case.
The Lie algebra $A_n$ acts naturally on $\hat{\E}_k$ and hence the symbols $\hat{g}_k$
are $A_n$-modules.

 \begin{proposition}\label{H^1T}
The defining equations of $\hat{\E}$ have second order. In other words, the PDEs of higher order
$k>2$ specifying $\hat{\E}_k$ are prolongations of the second-order PDEs.
  \end{proposition}

\smallskip

\noindent {\bf Proof.}
The statement is equivalent to the equality of symbolic prolongations:
$\hat{g}_k=\hat{g}_2^{(k-2)}$ for $k>2$. Here, similarly to the preceding proof we identify
$\hat{g}_k\subset S^kT^*$ where $T=T_o\Lambda$ is the model tangent space.
As is standard in the formal theory of differential equations \cite{Kras,KL}, the above equality of
prolongations is equivalent to the successive identities $\hat{g}_{k+1}=\hat{g}_k^{(1)}$, $k\ge2$,
and this is equivalent to the vanishing of the cohomology in the second non-trivial term
of the Spencer $\delta$-complex
 $$
0\to \hat{g}_{k+1}\stackrel{\delta}\longrightarrow T^*\ot \hat{g}_k\stackrel{\delta}\longrightarrow
\La^2T^*\ot \hat{g}_{k-1}\longrightarrow\dots
 $$
Namely $H^{1,k}(\hat{g})$ is the cohomology at the term $T^*\ot \hat{g}_k$. It is well-known \cite{Sp,Mal} that
the Spencer cohomology complex dualises over $\R$ to the Koszul homology complex
 \begin{equation}\label{Koszul}
0\leftarrow \hat{g}^*_{k+1}\stackrel{\ \partial}\longleftarrow
T\ot \hat{g}^*_k\stackrel{\ \partial}\longleftarrow
\La^2T\ot \hat{g}^*_{k-1}\longleftarrow\dots
 \end{equation}
Our claim is equivalent to the vanishing of the corresponding homology: $H_{1,k}(\hat{g}^*)=0$ for $k\ge2$.
From the preceding proof and \cite{MorMathOverflow} it follows that the symbols, considered as
$A_n$-modules, are
 $$
\hat{g}_k=S^kS^2V_n^*\cap S^2\La^kV_n^*=\Gamma_{2\lambda_{n-k+1}},
 $$
where $V_n=\Gamma_{\lambda_1}=\R^n$ is the standard representation
and $V_n^*=\Gamma_{\lambda_n}$ is its dual.
For $k=0$ we get $\hat{g}_0=\Gamma_0=\R$.
Dualising the symbol we get $\hat{g}_k^*=\Gamma_{2\lambda_k}$ (this is the main advantage:
the computations will be visibly $n$-independent);
for $k=0$ again $\hat{g}_0=\R$. Also $T=\Gamma_{2\lambda_1}$.

At this point we start working over $\C$: since the complexification does not change the rank
of the cohomology, this simplification does not restrict the generality.
The Littlewood-Richardson rule yields the following tensor decompositions for the second nonzero term
of the Koszul complex (it applies for $0<k\leq n$ and requires a modification otherwise):
 \begin{equation}\label{E1}
\Gamma_{2\lambda_1}\otimes\Gamma_{2\lambda_k}=\Gamma_{2\lambda_{k+1}}+
\Gamma_{\lambda_1+\lambda_k+\lambda_{k+1}}+\Gamma_{2\lambda_1+2\lambda_k}.
 \end{equation}
Similarly for the third nonzero term, using the plethysm $\La^2T=\Gamma_{2\lambda_1+\lambda_2}$,
we have for $k\ge2$
(the agreement is that $\lambda_0=\lambda_{n+1}=0$;
entries in parentheses below are equal for $k=2$ and add without multiplicity;
for $k=n+1$ the second and fifth terms disappear):
 \begin{multline}\label{E2}
\Gamma_{2\lambda_1+\lambda_2}\otimes\Gamma_{2\lambda_{k-1}}=
\Gamma_{\lambda_2+2\lambda_k}+\Gamma_{\lambda_1+\lambda_k+\lambda_{k+1}}+
(\Gamma_{\lambda_1+\lambda_2+\lambda_{k-1}+\lambda_k}+\Gamma_{2\lambda_1+2\lambda_k})\\
+\Gamma_{2\lambda_1+\lambda_{k-1}+\lambda_{k+1}}+
(\Gamma_{2\lambda_1+\lambda_2+2\lambda_{k-1}}+\Gamma_{3\lambda_1+\lambda_{k-1}+\lambda_k}).
 \end{multline}
Similarly, using the plethysm $\La^3T=\Gamma_{3\lambda_2}+\Gamma_{2\lambda1+\lambda_3}$,
we decompose the next term in \eqref{Koszul}:
 \begin{multline*}
\La^3T\ot\Gamma_{2\lambda_{k-2}}=
\Gamma_{\lambda_2+2\lambda_k}+\Gamma_{2\lambda_2+\lambda_{k-2}+\lambda_k}
+\Gamma_{3\lambda_2+2\lambda_{k-2}}+\Gamma_{\lambda_1+\lambda_3+2\lambda_{k-1}}
+2\Gamma_{\lambda_1+\lambda_2+\lambda_{k-1}+\lambda_k}\\
+\Gamma_{\lambda_1+2\lambda_2+\lambda_{k-2}+\lambda_{k-1}}
+\Gamma_{2\lambda_1+\lambda_{k-1}+\lambda_{k+1}}
+\Gamma_{2\lambda_1+\lambda_3+\lambda_{k-2}+\lambda_{k-1}}
+2\Gamma_{2\lambda_1+\lambda_2+2\lambda_{k-1}}
+\Gamma_{2\lambda_1+\lambda_2+\lambda_{k-2}+\lambda_k}\\
+\Gamma_{3\lambda_1+\lambda_{k-1}+\lambda_k}
+\Gamma_{3\lambda_1+\lambda_{k-2}+\lambda_{k+1}}
+\Gamma_{3\lambda_1+\lambda_3+2\lambda_{k-2}}
+\Gamma_{3\lambda_1+\lambda_2+\lambda_{k-2}+\lambda_{k-1}}
+\Gamma_{4\lambda_1+\lambda_{k-2}+\lambda_k}.
 \end{multline*}
Again, in the special cases $k\in\{2,3,n+1,n+2\}$ this decomposition requires a modification.
By Shur's lemma a homomorphism $\Gamma_\mu\to\Gamma_\nu$ is either zero or an isomorphism
in the case $\mu=\nu$. It can be checked that Young symmetrisers are nontrivial on the common terms
of \eqref{E1} and \eqref{E2}, i.e.\ the last two terms of \eqref{E1} are isomorphic images under
the Koszul differential $\partial$ of the same type modules from the decomposition \eqref{E2}.
The first term $\Gamma_{2\lambda_{k+1}}$ in the right-hand-side of \eqref{E1}
does not come however as the image of the second differential $\partial$ and it does not go to zero
under the first  differential $\partial$ in \eqref{Koszul}, i.e.\ it is mapped isomorphically to $g_{k+1}^*$.
Special cases $k=2,k=n$ have to be considered separately, but they lead to the same conclusion.

Thus the first Koszul homology vanishes, $H_{1,k}(\hat{g}^*)=0$, and by dualisation the first
Spencer cohomology does the same: $H^{1,k}(\hat{g})=0$, $k>1$. This implies the claim of the proposition.
 In the same vein, $H_{2,k}(\hat{g}^*)=0$ and hence $H^{2,k}(\hat{g}^*)=0$ for $k\ge2$.
\qed

As a corollary of this proposition we deduce the last building block for the main theorem of this section,
which will therefore finish the proof of Theorem \ref{T0}.

 \begin{lemma}\label{LeM}
For $k>2$ the following holds: $g_k=g_2^{(k-2)}$.
 \end{lemma}

\smallskip

\noindent {\bf Proof.}
Let us first note that the symbols of the two considered equations agree
except at degree one: $\hat{g}_k=g_k$ for $k\ge2$ (also $\hat{g}_0=\R=g_0$).
However they form symbolic complexes over different vector spaces: $T$ for $\hat{g}$
and $\tau$ for $g$. The relation between these spaces is given by the exact sequence
 $$
0\longrightarrow\tau\longrightarrow T\longrightarrow \R\longrightarrow0
 $$
that is induced by the embedding $X\hookrightarrow\Lambda$, identifying the normal bundle
with $\R$. Dualisation of this 3-sequence gives a relation between $\hat{g}_1=T^*$ and $g_1=\tau^*$;
it is used in the diagram below.
We unite the Spencer $\delta$-complexes for $\hat{g}$ and $g$ into the following commutative diagram
with exact vertical sequences:
 \begin{center}
\begin{tikzcd}
&& 0\arrow{d}& 0\arrow{d}& 0\arrow{d} \\
& 0\arrow{r}\arrow{d}& g_k\arrow{r}\arrow{d}& \tau^*\ot g_{k-1}\arrow{r}\arrow{d} &
\La^2\tau^*\ot g_{k-2}\arrow{r}\arrow{d} &\dots\\
0\arrow{r}& \hat{g}_{k+1}\arrow{r}\arrow{d}& T^*\ot \hat{g}_k\arrow{r}\arrow{d}& \La^2T^*\ot \hat{g}_{k-1}\arrow{r}\arrow{d} &
\La^3T^*\ot \hat{g}_{k-2}\arrow{r}\arrow{d} &\dots\\
0\arrow{r}& g_{k+1}\arrow{r}\arrow{d}& \tau^*\ot g_k\arrow{r}\arrow{d} &
\La^2\tau^*\ot g_{k-1}\arrow{r}\arrow{d} &
\La^3\tau^*\ot g_{k-2}\arrow{r}\arrow{d} &\dots\\
& 0 & 0& 0& 0&
\end{tikzcd}
 \end{center}
This diagram should be modified in the column with central term $\La^kT^*\ot\hat{g}_1$,
as the kernel of the projection map gets an additional factor, namely it becomes
$\La^{k-1}\tau^*\ot g_1\oplus\La^kT^*$.
By the standard diagram chase we know that $H^{1,k}(g)=H^{1,k-1}(g)$ provided
$H^{1,k}(\hat{g})=H^{2,k-1}(\hat{g})=0$.
This and the result of Proposition \ref{H^1T} imply the following relations:
 $$
H^{1,1}(g)\oplus\La^2T^*\supset H^{1,2}(g)=H^{1,3}(g)=\dots
 $$
Since $H^{1,n+2}(g)=0$ 
by dimensional reasons, the above sequence stabilises at zero
and hence $H^{1,k}(g)=0$ for $k\ge2$. Vanishing of this Spencer cohomology
is equivalent to the equality $g_{k+1}=g_k^{(1)}$, and the result follows.
\qed

\subsection{Monge-Amp\`ere equations in implicit form}\label{sec:MAEs}

In implicit form, symplectic Monge-Amp\`ere equations can be expressed as  linear relations among minors of
the Hessian matrix $U$, namely $F=0$ with
 \begin{equation}\label{sMAEs}
F=\sum_\sigma a_\sigma\det U_\sigma,
 \end{equation}
where $\sigma$ encodes minors of a symmetric $(n+1)\times(n+1)$ matrix of any size $|\sigma|\in[0,n+1]$.
Note that there are relations among minors starting from $n=3$, so a basis of minors
should be chosen. Equations of this form can be uniquely characterised by their defining relations.

 \begin{theorem}\label{xMAE}
The following linear system $\hat\E$ of PDEs  of the second order,
 \begin{gather*}
F_{u_{ii}u_{ii}}=0,\ \qquad F_{u_{ii}u_{ij}}=0,\ \qquad 2F_{u_{ii}u_{jj}}+F_{u_{ij}u_{ij}}=0,\\
F_{u_{ij}u_{kk}}+F_{u_{ik}u_{jk}}=0,\ \quad \ F_{u_{ij}u_{kl}}+F_{u_{ik}u_{lj}}+F_{u_{il}u_{jk}}=0,
 \end{gather*}
is involutive. Its solution space is generated by Monge-Amp\`ere equations \eqref{sMAEs} (all indices $i,j,k,l$ run from $0$ to $n$ and are assumed pairwise distinct).
 \end{theorem}

\smallskip

\noindent  {\bf Proof.}
An elementary approach to obtain this system is to calculate second-order partial derivatives of the function $F$ defined by \eqref{sMAEs}  and to eliminate the parameters $a_\sigma$.
This works well for small $n\leq 4$ and gives the required relations
(in low dimensions one can  verify directly that
the defining relations for $F$ of order $>2$ are prolongations of the relations of order $2$).

The necessity of these equations follows from the fact that every traveling wave reduction of a Monge-Amp\`ere equation
is a Monge-Amp\`ere equation in lower dimensions, and the reduction to $\R^4(x_i,x_j,x_k,x_l)$
yields the claim.

As for the sufficiency (recall that  $u$ is a function of $n+1$ arguments), let us first note that the number of equations in the system
is $\hat{N}(n+1)$ where
$\hat{N}(n)=n+n(n-1)+\binom{n}{2}+n\binom{n-1}{2}+\binom{n}{4}=\binom{n+3}{4}$,
so the 2-symbol of $\hat{\E}$ has dimension
$\binom{d(n+1)+1}{2}-\hat{N}(n+1)=\frac{n(n+1)^2(n+2)}{12}$.
Since this is equal to the dimension of the $A_n$-module $\Gamma_{2\lambda_2}$,
namely $\dim\Gamma_{2\lambda_2}=\dim S^2\La^2T-\dim\La^4T=\frac{n+1}2\binom{n+2}{3}$
for $\dim T=n+1$,
it follows that the symbol space coincides with $\hat{g}_2$ as discussed in the previous section
(the fact that $\hat{g}_2$ is a subspace of the 2-symbol of $\hat{\E}$ follows from the necessity
of the above relations).

This in turn implies that $\hat{\E}$ coincides with the equation (also denoted $\hat {\E}$)
from the previous section (note that they are formally different: the equation from the previous section
is defined by jets of its solutions, while $\hat{\E}$ from this section is given by explicit linear relations
and their prolongations). Indeed, this equation is involutive by Proposition \ref{H^1T}, in particular
the first Spencer cohomology vanishes, $H^{1,k}(\hat{\E})=H^{1,k}(\hat{g})=0$ for $k>1$,
meaning that all equations of order $k+2$ in the system describing $\hat{\E}$ are obtained by
$k$-differentiations of the equation given in the formulation of the theorem.
This finishes the proof.
 \qed

\noindent {\bf Remark:} Complex \eqref{Koszul} yields that
$H^{1,1}(\hat{g})=\Gamma_{4\lambda_1}$ is an irreducible module. In particular,
all equations for $F$ in Theorem \ref{xMAE} follow from any one of them via $A_n$-representation,
cf.\ the Remark at the end of Section \ref{sec:MA3}.

\medskip

\noindent {\bf Remark:} There is yet another form of characterising symplectic Monge-Amp\`ere
property by the system whose solutions have the same locus $F=0$ in $J^2(\R^{n+1})$ as \eqref{sMAEs}.
Namely, if we allow arbitrary reparametrisations $F\mapsto \Psi(F)$ for a diffeomorphism $\Psi\in\op{Diff}(\R)$,
the equation takes the form $F=\op{const}$. Making this substitution in the system of Theorem \ref{xMAE}
in 2D gives the following nonlinear system (where we again use the  notation
$r=u_{xx},\ s=u_{xy},\ t=u_{yy}$):
 \begin{gather*}
F_{rr}=\frac{F_r^2}{2F_rF_t+F_s^2}(2F_{rt}+F_{ss}),\quad
F_{rs}=\frac{F_rF_s}{2F_rF_t+F_s^2}(2F_{rt}+F_{ss}),\\
F_{tt}=\frac{F_t^2}{2F_rF_t+F_s^2}(2F_{rt}+F_{ss}),\quad
F_{st}=\frac{F_sF_t}{2F_rF_t+F_s^2}(2F_{rt}+F_{ss}).
 \end{gather*}
In higher-dimensional cases the situation is similar, but equations become somewhat more cumbersome.
The resulting system  is of infinite type, with the general solution depending on an arbitrary function of
one argument. In 2D  the general solution of the above system is
 $$
F=\Psi(a_0+a_1r+a_2s+a_3t+a_4(rt-s^2))
 $$
where the constants $a_i$ and the function $\Psi$ are arbitrary.

Since the pseudogroup $G=\op{Diff}_\text{loc}(\R)$
acts on the above system of PDEs, it is possible to compute the quotient.
One way to do this is to derive  scalar differential invariants and rational syzygies between them
as explained in \cite{KL-GLT}. Another possibility is to fix a cross-section of the orbit foliation
by choosing a proper normalisation. Making the solution either polynomial of order $n+1$
(this defines $F$ up to multiplication by a nonzero constant) or expressing $F=0$ in the evolutionary
form $u_{00}=f$ we obtain determining equations $\hat{\E}$ or $\E$, respectively.

\section{Integrability and geometry of Hirota type equations}\label{sec:Proofs}

After a few remarks on the action of the equivalence group
we prove the main result about Hirota type equations in 4D. All calculations
are based on computer algebra systems \textsf{Mathematica} and \textsf{Maple}
(these only utilise symbolic polynomial algebra over $\mathbb{Q}$, so the results are rigorous). The programmes are available from the arXiv version of this paper.
At the end of this section we use the reduction argument to conclude that integrability in dimensions higher
than four also implies the Monge-Amp\`ere property.

\subsection{Action of the equivalence group}\label{sec:LieGrpSp8}

The group $G={\bf Sp}(2n+2,\C)$ acts naturally  on $\C^{2n+2}=T^*(\C^{n+1})$, with the stabiliser of a
Lagrangian plane $\C^{n+1}\subset\C^{2n+2}$ being a parabolic subgroup $P$ corresponding to the Dynkin diagram
 \[
 \begin{tikzpicture}
\bond{0,0}; \bond{1,0}; \bond{3,0}; \lbond{4,0};
\wDnode{0,0}; \wDnode{1,0}; \wDnode{4,0}; \xDnode{5,0};
\path[dotted](2,0) edge (3,0);
 \end{tikzpicture}
 \]
Here the cross indicates the parabolic subgroup. This leads to the homogeneous representation of the (complex)
Lagrangian Grassmannian $\Lambda=G/P$. The stabiliser of a point $o\in\Lambda$ is $P$,
and it acts on jets of hypersurfaces $X$ through this point. In particular, $G$ acts on the space
$J^k_1(\Lambda)$ of $k$-jets of codimension 1 submanifolds $X\subset\Lambda$
(an affine chart of this is the standard jet-space $J^k\bigl(\C^{\bar{d}}\bigr)$) and $P$ acts on the space $J^k_1(\Lambda)_o$ of $k$-jets of codimension 1 submanifolds through $o$.

This action on $J^1_1(\Lambda)$ has a unique open orbit corresponding
to 1-jets of  non-degenerate hypersurfaces $X$. Indeed, this is equivalent to the uniqueness
of non-degenerate linear second-order equations up to complex symplectic transformations
(over $\R$ there are $[\frac{n+1}2]+1$ open orbits). This justifies our computational trick described 
in the introduction, namely that we can evaluate any
function on $J^k_1(\Lambda)$  by restricting it to the fibre over a non-degenerate 1-jet.

We claim that transformations from the equivalence group $G$ 
correspond to special contact transformations from a different jet-space $J^2(\R^{n+1})$ that naturally
act on equations of type \eqref{H}, cf.\ \cite{Fer4, DF}
(at this point we are not concerned with the classification and switch to the real case).
Let us clarify this in the affine chart $U\in S^2\R^{n+1}\subset\Lambda$,
where the generators of the Lie algebra $\mathfrak{g}=\op{Lie}(G)=\mathfrak{sp}(2n+2)$
are $X_{\alpha\beta},L_{\alpha\beta},P_{\alpha\beta}$ as in Section \ref{sec:Sp}.
Consider the space of 2-jets of functions $u=u(x^0,x^1,\dots,x^n)$, namely
$J^2(\R^{n+1})\simeq\R^{n+1+d(n+2)}$ with coordinates $(x^i,u,u_j,u_{ij})$.
The algebra $\mathfrak{g}$ acts naturally
on $T^*(\R^{n+1})\simeq\R^{2n+2}(x^i,u_j)$ by linear symplectic transformations, so it is contained in the
algebra of contact vector fields in $J^2$.
Indeed,  on restriction to fibres of the bundle $J^2(\R^{n+1})\to J^1(\R^{n+1})$, 
prolongations of the point vector fields
$\xi_{\alpha\beta}=\frac1{1+\delta_{\alpha\beta}}\,x^\alpha x^\beta\partial_u$,
$\eta_{\alpha\beta}=-x^\alpha\partial_{x^\beta}$ and the contact vector fields
$\zeta_{\alpha\beta}=-u_\alpha\partial_{x^\beta}-u_\beta\partial_{x^\alpha}-u_\alpha u_\beta\partial_u$
coincide with the vector fields $X_{\alpha\beta}$, $L_{\alpha\beta}$ and $P_{\alpha\beta}$, respectively.
Thus, $\xi_{\alpha\beta}^{(2)}F(U)=X_{\alpha\beta}F(U)$, $\eta_{\alpha\beta}^{(2)}F(U)=L_{\alpha\beta}F(U)$
and $\zeta_{\alpha\beta}^{(2)}F(U)=P_{\alpha\beta}F(U)$ for all functions $F=F(U)$ of type \eqref{H}.

\subsection{Integrability: proof of Theorem \ref{T1}}\label{sec:Proof-2}

In this and the next two sections we restrict to the four-dimensional case ($n=3$).

\medskip

\noindent {\bf Equivalence $(a)\Longleftrightarrow (d)$.}

\noindent Here we apply the method of hydrodynamic reductions to a general Hirota type equation written in  evolutionary form (\ref{m0}).   Our strategy is to derive a set of constraints for the right-hand side $f$ that are necessary and sufficient for  integrability.
As outlined in  \cite{Fer4}, in 3D this leads to an involutive system of third-order differential constraints for $f$.
The crucial difference occurring in the 4D case is the appearance, along with third-order differential constraints, of additional second-order integrability conditions that  imply the Monge-Amp\`ere property.  The rest follows from the classification of integrable symplectic Monge-Amp\`ere equations in 4D \cite{DF}. Thus,  the requirement of  integrability in 4D is far more rigid than that in 3D.
 
The proof of the implication $(a)\implies(d)$ is as follows. Representing Hirota type equation in evolutionary form  (\ref{m0})$_{n=3}$
and introducing the notations
\begin{gather*}
u_{01}=d,\ u_{02}=r,\ u_{03}=n,\ u_{11}=a,\ u_{12}=b,\ u_{13}=c,\ u_{22}=p,\ u_{23}=q,\ u_{33}=m,\\
u_{00}=f(d, r, n, a, b, c, p, q, m),
\end{gather*}
we transform (\ref{m0})$_{n=3}$ into quasilinear form (\ref{quasi1})$_{n=3}$ by adding
the compatibility conditions $(u_{\alpha\beta})_\gamma=(u_{\alpha\gamma})_\beta$, i.e.
 $$
d_{x^0}=f_{x^1},\  d_{x^1}=a_{x^0},\  d_{x^2}=b_{x^0}, \ d_{x^3}=c_{x^0}, \ \text{etc.}
 $$
Ansatz (\ref{ansatz}) requires  that  the new dependent variables $d, r, n, a, b, c, p, q, m$ are sought as functions of the  phases $R^{I},\ I=1, \dots N$, which themselves satisfy a triple of hydrodynamic type systems (\ref{R}).
 This implies the relations
 \begin{equation}
 \begin{array}{c}
\partial_Ib=\mu^I\partial_Ia, \ \partial_Ic=\nu^I\partial_Ia, \ \partial_Id=\lambda^I\partial_Ia, \\
\ \\
 \partial_Ir=\lambda^I \mu^I \partial_Ia, \ \partial_In=\lambda^I \nu^I \partial_Ia, \ \partial_Iq=\mu^I\nu^I \partial_Ia, \ \partial_Ip=(\mu^I)^2 \partial_Ia,\ \partial_Im=(\nu^I)^2 \partial_Ia,
 \end{array}
 \label{relat}
 \end{equation}
$\partial_I=\partial_{R^I}$,  no summation assumed. Furthermore, the characteristic speeds
 $\mu^I, \nu^I, \lambda^I$ must satisfy the dispersion relation,
 \begin{equation}
 \begin{array}{c}
( \lambda^I)^2=f_a+f_b\mu^I+f_c\nu^I+f_d\lambda^I
+f_r\lambda^I \mu^I +f_n\lambda^I \nu^I +f_q\mu^I\nu^I+f_p(\mu^I)^2 +f_m(\nu^I)^2 .
\end{array}
 \label{DR}
 \end{equation}
 Differentiating the dispersion relation by $R^J, \ J\ne I,$ we obtain
 \begin{equation}
 \frac{\partial_J\mu^I}{\mu^J-\mu^I}=\frac{\partial_J\nu^I}{\nu^J-\nu^I}=\frac{\partial_J\lambda^I}{\lambda^J-\lambda^I}=B_{IJ}\partial_Ja,
 \label{BIJ}
 \end{equation}
(no summation) where $B_{IJ}$ are  rational expressions in $\mu^I, \mu^J,  \nu^I, \nu^J, \lambda^I, \lambda^J$ whose coefficients depend on partial derivatives of  $f$ up to the second order (we do not present them here explicitly). Calculating consistency conditions for relations (\ref{relat}) we obtain  the symmetry condition $B_{IJ}=B_{JI}$ (which is satisfied identically), as well as the following equations for $a$:
\begin{equation}
\partial_I\partial_Ja=2B_{IJ}\partial_Ia\partial_Ja.
 \label{aIJ}
\end{equation}
 Finally,  the consistency conditions for relations (\ref{BIJ}) and (\ref{aIJ}) take the form
 \begin{equation}
 \partial_KB_{IJ}=(B_{IK}B_{JK}-B_{IK}B_{IJ}-B_{IJ}B_{JK})\partial_Ka,
 \label{fc}
 \end{equation}
$I\ne J \ne K$ (without any loss of generality one can set $I=1, J=2, K=3$). Calculating the left-hand side of (\ref{fc}) via (\ref{relat}), (\ref{BIJ}), (\ref{aIJ}), and  utilising the dispersion relation (\ref{DR}) to eliminate all higher powers of $\lambda^I$, one can reduce (\ref{fc}) to  a  polynomial expression in $\mu^I, \mu^J, \mu^K, \nu^I, \nu^J, \nu^K$, which also depends linearly on $\lambda^I, \lambda^J, \lambda^K$;
note that the common factor $\partial_Ka$ will cancel. Equating to zero the coefficients of this polynomial
we obtain a system $S$ of third-order PDEs for $f$ -- the required integrability conditions.
This system will be linear in the third-order partial derivatives of $f$, and quadratic in the second-order derivatives.

In the 3D case system $S$  can be uniquely solved for all of the 35 third-order partial derivatives of $f$, resulting in the  35 integrability conditions that are in involution: there will be no second-order
relations left \cite{Fer4}.
The {\bf first remarkable phenomenon} of the 4D case is that, after solving system $S$ for all  of the 165 third-order partial derivatives of $f$, there will still be numerous homogeneous quadratic relations in the second-order derivatives of $f$ remaining (over 2000 quadratic relations).
The {\bf second remarkable phenomenon} is that the radical of  the ideal  generated by these quadratic relations contains  all of the 25 linear (in 2-jets of $f$) relations characterising Monge-Ampere systems in 4D, see Section \ref{sec:MAh}.
This establishes the Monge-Amp\`ere property, and thus finishes the proof of the implication
$(a)\implies (d)$ due to the existing classification of integrable symplectic Monge-Amp\`ere equations in 4D \cite{DF}.
Let us note that the computation of the radical of the quadratic ideal simplifies dramatically if one gives the first-order derivatives of $f$ some generic constant numerical values, see Section \ref{sec:Sp} for the discussion. We have chosen $f_b=f_n=1$,  all other $f_i=0$.

The converse implication, $(d)\implies (a)$, is a straightforward  calculation based on normal  forms of symplectic Monge-Amp\`ere equations from Theorem \ref{SMA}, see e.g. \cite{Fer2, Fer3}.
\medskip

\noindent {\bf Equivalence $(b)\Longleftrightarrow (d)$.}

\noindent The proof of  the implication $(b)\implies (d)$ is based on a direct computation of the Weyl tensor of the conformal structure $[g]$. First we demonstrate that either of the  half-flatness conditions, $W_-=0$ or $W_+=0$, implies that the 4D equation under study must be of symplectic Monge-Amp\`ere type. Here  $W_{-}=\frac{1}{2}(W-*W), \
W_{+}=\frac{1}{2}(W+*W)$.
As above we use the 1-jet of  $f$ defined as
$f_b=f_n=1$, all other $f_i=0$. Let us substitute this 1-jet into one of the half-flatness conditions, say  $W_-=0$, reduce it modulo (\ref{m0}), and equate to zero coefficients  at the fourth-order derivatives  $u_{ijkl}$.  This will give a  linear system in the
2-jet of $f$ (30 linear relations altogether). A direct verification shows that this linear system implies all of the 25 conditions characterising
Monge-Amp\`ere equations in 4D (in which we substitute  the same 1-jet of $f$). For $W_+=0$  considerations are essentially the same. Thus, we have established the Monge-Amp\`ere property.

Furthermore, due to the half-flatness of $[g]$, equation (\ref{H}) possesses a dispersionless Lax pair \cite{CalKrug}. Thus, any travelling wave reduction of this  equation to 3D  is a  symplectic Monge-Amp\`ere equation possessing a Lax pair; hence, the reduction  must be linearisable. The rest follows from the classification of integrable symplectic Monge-Amp\`ere equations in 4D possessing linearisable travelling wave reductions \cite{DF}.
Let us note that the half-flatness conditions, $W_+=0$ and $W_-=0$,  are not equivalent: in fact, only one of them leads to integrable  Monge-Amp\`ere equations, while the other one  is much more overdetermined, and  implies linearisability. There is however no invariant way to distinguish between them
(due to the lack of a canonically defined orientation),  so we  just state that conformal
half-flatness implies the Monge-Amp\`ere integrability.

The converse implication, $(d)\implies (b)$, is a straightforward computation based on normal  forms of symplectic Monge-Amp\`ere equations from Theorem \ref{SMA}: it was explicitly noted in Section 8 of \cite{FerKrug}.
\medskip

\noindent {\bf Equivalence $(b)\Longleftrightarrow (c)$.}

\noindent This is a particular case of the general result of \cite{CalKrug} relating self-duality of
the conformal structure $[g]$ to the existence of a dispersionless Lax pair.

Let us give a few more details on the implication $(c)\Longrightarrow (b)$. The main technical
result of \cite{CalKrug} is that any nontrivial Lax pair must be characteristic, i.e.\ null with respect
to the conformal structure. For every solution $u$ of \eqref{H} the congruence of null two-planes defined by the Lax pair
uniquely lifts into the correspondence space $\hat{M}_u\to M_u$, where
$M_u=\op{graph}(u)\subset\R^5(x^1,x^2,x^3,x^4,u)$ and
$\hat{M}_u\simeq M\times\mathbb{P}(\lambda)$ is the bundle of null $\alpha$-planes (self-dual 2-planes).
The corresponding 2-distribution in $\hat{M}_u$ is Frobenius-integrable
for every solution $u$, and thus by projection we obtain a 3-parameter family of $\alpha$-surfaces,
i.e.\ totally null  surfaces of the conformal structure $[g]$ on $M_u$.
According to Penrose \cite{Penrose} this is equivalent to self-duality.


The converse implication is easily seen by transitivity,  $(b)\Rightarrow(d)\Rightarrow(c)$. Indeed,
since $(b)\implies(d)$ is already established, the claim follows from the fact that all equations from
Theorem \ref{SMA} are known to possess dispersionless  Lax pairs, see e.g.\ \cite{DF}.

This finishes the proof of Theorem \ref{T1}. \qed

\subsection{Linearisability: proof of Theorem \ref{T2}}\label{sec:linearisation}

It is clear that if a second-order PDE is linearisable (more precisely, transformable to a constant-coefficient linear form) by a  contact transformation, then
the corresponding conformal structure is flat on every solution.
Conversely, suppose that the Weyl tensor $W$ vanishes  on every solution.
Then also $W_-=0$, so the  conformal structure is self-dual on every solution.
Therefore,  by Theorem \ref{T1} the equation must be integrable, and  of symplectic Monge-Amp\`ere type.
Moreover, up to a transformation from the equivalence group ${\bf Sp}(8)$
it reduces to one of the six normal forms from Theorem \ref{SMA}. A straightforward computation shows that
$W\not\equiv0$ on a generic solution for the last five equations from the list.
Thus, the equation must be of the first type, and hence linearisable.
\medskip

\noindent {\bf Corollary:}
{\it Hirota type equation (\ref{H}) is linearisable by a transformation from the equivalence group ${\bf Sp}(8)$
if and only if it is linearisable by a contact transformation.}

\subsection{Symmetry: proof of Theorem \ref{T3}}\label{symalg}

 For each of the integrable
symplectic Monge-Amp\`ere equations from Theorem \ref{SMA}, their symmetry algebras $\mathfrak{s}$ inside $\mathfrak{sp}(8)$ were computed  in \cite{DF}.
The full (infinite-dimensional) contact symmetry algebras $\mathfrak{sym}$
of the same equations were computed in \cite{KrugMor2}.
However, in neither of these references the Lie algebra structure of
$\mathfrak{s}=\mathfrak{sym}\cap\mathfrak{sp}(8)$ was investigated. Here we fill this gap by a straightforward
application of  the {\sl LieAlgebras} package of {\sf Maple} and the standard Lie theory.

The following Table summarises the results. Note that we have changed the linear hyperbolic equation
from the list of Theorem \ref{SMA}
to the  ultra-hyperbolic form $u_{00}+u_{11}-u_{22}-u_{33}=0$ with the
 conformal structure of signature (2,2): for this signature the null planes are real.
Though the classification in Theorem \ref{SMA} is over $\mathbb{C}$, we provide finer
Lie algebra structures over $\mathbb{R}$, writing $\mathfrak{sl}_2=\mathfrak{sl}(2,\R)$ and so on.
The results over $\C$ are obtained by complexification using
 $\mathfrak{so}(2,2)^\C=\mathfrak{s0}(4,\C)$,
$\mathfrak{sl}(2,\C)^\C=\mathfrak{sl}_2(\C)\oplus\mathfrak{sl}_2(\C)$, etc.

 \begin{center}
 \begin{tabular}{ | l | c | l |} \hline
{\footnotesize Equation} & {\footnotesize\!dim($\mathfrak{s}$)\!}  & {\footnotesize Levi decomposition of $\mathfrak{s}$} \\
 \hline
{\footnotesize Linear ultrahyperbolic}&
16 &
{\footnotesize $\mathfrak{s}=\mathfrak{so}(2,2)\ltimes S^2\R^4\simeq\mathfrak{co}(2,2)\ltimes S^2_0\R^4$} \\
\hline
{\footnotesize Second heavenly}&
14 &
\begin{minipage}{10cm}{\hphantom{.}\footnotesize $\mathfrak{s}=\mathfrak{sl}_2\ltimes\mathfrak{rad}$;
$\mathfrak{rad}=\R^2+V_1+V_2+V_3$ as $\mathfrak{sl}_2$-module\\
($\R^2$ is a trivial module; $V_i$ are 3D irreps);\\
As Lie algebra: $\R^2=\mathfrak{sol}_2=\langle s,t:[s,t]=t\rangle$,\\
$\op{ad}_s|V_i=i$, $\op{ad}_t(V_i)=V_{i+1}$, $[V_i,V_j]=V_{i+j}$}
\vphantom{$\frac{a}{a}$}\end{minipage}\\
\hline
{\footnotesize Modified heavenly} &
13 &
\begin{minipage}{10cm}{\hphantom{.}\footnotesize $\mathfrak{s}=\mathfrak{sl}_2\oplus
(\mathfrak{sl}_2\ltimes\mathfrak{rad})$, where
$\mathfrak{rad}=\R+V_1+V_2$ as $\mathfrak{sl}_2$-module\\
($\R$ is a trivial module; $V_i$ are 3D irreps);\\
As Lie algebra: $\R=\langle s\rangle$,
$\op{ad}_s|V_i=i$, $[V_i,V_j]=V_{i+j}$}
\vphantom{$\frac{a}{a}$}\end{minipage}\\
\hline
{\footnotesize First heavenly}&
13 &
\begin{minipage}{10cm}{\hphantom{.}\footnotesize $\mathfrak{s}=(\mathfrak{sl}_2\oplus\mathfrak{sl}_2)
\ltimes\mathfrak{rad}$, where
$\mathfrak{rad}=\R+V_1+V_2$ as $\mathfrak{sl}_2^2$-module\\
($\R$ is trivial; $V_1/V_2$ are 3D irreps of the first/second copy of $\mathfrak{sl}_2$);\\
As Lie algebra: $\R=\langle s\rangle$, $\op{ad}_s|V_i=(-1)^i$, $[V_i,V_j]=0$}
\vphantom{$\frac{a}{a}$}\end{minipage}\\
\hline
{\footnotesize Husain equation} &
12 &
\begin{minipage}{10cm}{\footnotesize $\mathfrak{s}=\mathfrak{sl}_2(\C)\oplus
(\mathfrak{sl}_2\ltimes V)$ where $V$ is a 3D irrep}\end{minipage}\\
\hline
{\footnotesize General heavenly} &
12 &
\begin{minipage}{10cm}{\footnotesize $\mathfrak{s}=\mathfrak{sl}_2\oplus\mathfrak{sl}_2\oplus
\mathfrak{sl}_2\oplus\mathfrak{sl}_2$}\end{minipage}\\
\hline
\end{tabular}
 \end{center}

\medskip
It is apparent from the Table that the symmetry algebra of
every equation contains $\mathfrak{sl}_2$, and that the minimal dimension of the symmetry algebra is 12.
 It was shown in \cite{DF} that each equation-manifold $X^9\subset\Lambda^{10}$ contains
a  subvariety $X^4$ along which $X^9$ is singular.
Thus,  $X^4$ is intrinsic to the problem and so is invariant under the symmetry group $\mathcal{S}$.
Therefore the action of $\mathcal{S}$ on $X^9$ is not transitive.

\smallskip

\noindent {\bf Proof of Theorem \ref{T3}.}  Since the symmetry generators are known explicitely,
it is straightforward to verify that the action of the symmetry algebra $\mathfrak{s}$ has an open orbit.
As the action is algebraic and $X$ is irreducible, it is a Zariski open orbit. Moreover,
 singular orbits form an algebraic stratified submanifold of $X^9$ of positive codimension.
Since such submanifolds do not separate domains in $X^9$, there is precisely one Zariski open orbit of the
symmetry group $\mathcal{S}$. \qed

Two remarks about this proof are in order. First, the set of singular orbits is strictly bigger than
the singular variety $X^4\subset X^9$. Second, in the complex case the unique Zariski open orbit is connected in the usual
topology. However, when doing classification over $\R$
(in this case the classification from Theorem \ref{SMA} will contain more normal forms),
the set of regular points can be topologically disconnected.
For instance, for the modified heavenly equation the rank of  vector fields from $\mathfrak{s}$
drops to $8$ on the hypersurface $\{u_{03}=0\}\subset X^9$, and this hypersurface separates $X^9$
into two open pieces (in the set-theoretic topology).

\subsection{Integrability in higher dimensions: proof of Theorem \ref{T4}}\label{final-higherD}

The Monge-Amp\`ere property of higher-dimensional integrable Hirota type equations is a direct consequence of the analogous result in 4D.
First of all, a generic 4D traveling wave reduction of a multi-dimensional integrable non-degenerate Hirota type equation will also be non-degenerate and integrable, and hence of the symplectic Monge-Amp\`ere type by Theorem \ref{T1}. In particular,
all generic traveling wave reductions to 2D will  be of the symplectic Monge-Amp\`ere type. Thus, as noted in the Remark at the end of Section \ref{sec:MA3} (and also in the Remark at the end of Section \ref{sec:MAEs}), 
the equation will satisfy all relations (\ref{E1ab})-(\ref{E4}), and therefore will itself be of the symplectic Monge-Amp\`ere type by Theorem \ref{T0}. \qed

\medskip

\noindent {\bf Remark:} Although the classification of higher-dimensions integrable equations is still open, the lack of non-trivial  examples makes it tempting to conjecture that all multi-dimensional (5D and higher) {\it non-degenerate} integrable Hirota type equations  must be linearisable. We emphasize that the well-known integrable  6D version
of the second heavenly equation \cite{Takasaki1, Przanovski},
 $$
u_{15}+u_{26}+u_{13}u_{24}-u_{14}u_{23}=0,
 $$
does not constitute a counterexample to this conjecture: 
the corresponding symmetric bi-vector 
$\partial_1\partial_5+\partial_2\partial_6+u_{13}\partial_2\partial_4+u_{24}\partial_1\partial_3
-u_{14}\partial_2\partial_3-u_{23}\partial_1\partial_4$ has rank 4, and 
therefore the characteristic variety of the equation (and hence the equation itself) is degenerate.

\com{
\section{Concluding remarks}

This paper completes the classification of integrable Hirota type equations in 4D: all of them must be of symplectic Monge-Amp\`ere type, and are ${\bf Sp}(8)$-equivalent to one of the 6 normal forms from Theorem \ref{SMA}. Although the situation in higher dimensions is still open, the lack of non-trivial examples makes it tempting to conjecture that all multidimensional (5D and higher) {\it non-degenerate} integrable Hirota type equations  must be linearisable. We emphasize that the well-known integrable  6D version
of the second heavenly equation \cite{Takasaki1, Przanovski},
$$
u_{15}+u_{26}+u_{13}u_{24}-u_{14}u_{23}=0,
$$
does not constitute a counterexample to this conjecture:  its characteristic variety defines a quadratic form of rank 4,  therefore, the equation is degenerate. We hope to return to the general higher-dimensional case elsewhere.
}

\section*{Acknowledgements}

We thank B. Doubrov and M. Pavlov for clarifying discussions.
We acknowledge the enlightening discussion in MathOverflow initiated by G. Moreno \cite{MorMathOverflow};
in particular we exploit the formula conjectured by R. Bryant and proved by
D. Alekseevsky, G. Moreno, and independently by D.~Speyer.
This research was  supported by the EPSRC grant  EP/N031369/1.


\end{document}